\documentclass[preprint]{imsart}

\RequirePackage[colorlinks,citecolor=blue,urlcolor=blue]{hyperref}

\usepackage{amsmath}
\usepackage{amssymb}
\usepackage{graphicx}
\usepackage{natbib}

\usepackage[]{lineno}

\startlocaldefs
\newcommand{\beq}{%
\ifLineNumbers \begin{linenomath}\begin{equation}
\else \begin{equation} \fi}
\newcommand{\eeq}{%
\ifLineNumbers \end{equation}\end{linenomath}
\else \end{equation} \fi}
\newcommand{\bea}{%
\ifLineNumbers \begin{linenomath}\begin{eqnarray}
\else \begin{eqnarray} \fi}
\newcommand{\eea}{%
\ifLineNumbers \end{eqnarray}\end{linenomath}
\else \end{eqnarray} \fi}
\newcommand{\bes}{%
\ifLineNumbers \begin{linenomath}\begin{subequations}\begin{eqnarray}
\else \begin{subequations}\begin{eqnarray} \fi}
\newcommand{\ees}{%
\ifLineNumbers \end{eqnarray}\end{subequations}\end{linenomath}
\else \end{eqnarray}\end{subequations} \fi}
\newcommand{\mbf}[1]{\mathbf{#1}}

\newcommand{\msf}[1]{\mathsf{#1}}
\newcommand{\del}{\nabla}
\newcommand{\dsub}[1]{\partial_{#1}}
\newcommand{\given}{\, \vert \,}

\newcommand{\abs}[1]{{\vert {#1} \vert}}

\newcommand{\mean}[2]{{\langle {#1} \rangle}_{#2}}
\newcommand{\minfty}{{-\infty}}

\newcommand{\bigq}[2]{Q^{#1}_{#2}}
\newcommand{\p}[2]{p^{#1}_{#2}}
\newcommand{\q}[2]{q^{#1}_{#2}}
\endlocaldefs

\begin{document}

\begin{frontmatter}

\title{Applications of the Beta Distribution Part 1: Transformation Group Approach}
\runtitle{Applications of the Beta Distribution Part 1}


\author{\fnms{Robert W.} \snm{Johnson}\ead[label=e1]{robjohnson@alphawaveresearch.com}}
\address{29 Stanebrook Ct., Jonesboro, GA 30238 \\ \printead{e1}}
\affiliation{Alphawave Research}

\runauthor{Robert W. Johnson}

\begin{abstract}
A transformation group approach to the prior for the parameters of the beta distribution is suggested which accounts for finite sets of data by imposing a limit to the range of parameter values under consideration.  The relationship between the beta distribution and the Poisson and gamma distributions in the continuum is explored, with an emphasis on the decomposition of the model into separate estimates for size and shape.  Use of the beta distribution in classification and prediction problems is discussed, and the effect of the prior on the analysis of some well known examples from statistical genetics is examined.
\end{abstract}



\end{frontmatter}



\section{Introduction}

The beta distribution of the first kind, usually written in terms of the incomplete beta function, can be used to model the distribution of measurements whose values all lie between zero and one.  It can also be used to model the distribution for the probability of occurrence of some discrete event.  The most widely known technique for estimating the parameters, the method of moments, simply selects that beta distribution with the same first and second moments as found empirically from the data.  However, such a procedure is not well-justified from the perspective of probability theory.  To evaluate the reliability of the estimate of a model's parameters, as well as to determine the net evidence for a particular model relative to some other, one needs to follow the mathematical procedure which has come to be known as Bayesian data analysis.

Use of the beta distribution can be found in a variety of applications; for an overview of this and related classes of discrete statistical models and their use in Bayesian analysis, see \citet{pereira-2008special}.  One common use is as a model for an input process within a stochastic simulation \citep{kuhl-1747}.  Another is in the calculation of costs expected from a civil or industrial engineering project \citep{betk-179}.  It also has widespread use in the study of population genomics \citep{balding-963,price-ng1847}.  This paper concerns itself not so much with the choice of application but rather focuses on the methodology used to evaluate the parameters of the model given a set of measurements and the relative merit of competing models.  Various methods have been suggested for the estimation of its parameters, including the method of moments~\citep{abourizk-589} and variants of the Kolmogorov-Smirnov test~\citep{Press-1992}, as well as tests based on Bayesian significance values~\citep{pereira-2008can,bernardo-2012AIPC,Stern-11082013}. However, in this paper we will follow the traditional approach based on Bayes factors expressed in terms of the joint distribution for the quantities of interest.

This paper is organized as follows.  After a brief description of Bayesian data analysis, we explore the relation between the beta distribution and the Poisson and gamma distributions in the continuum.  The joint density for the size and shape parameter estimates can be expressed in alternate coordinate systems through geometric transformations which preserve the volume.  Marginalization over the size parameter leaves behind the beta distribution which describes the shape (distributed occurrence of events) of the possible outcomes.  After that, we examine the use of the beta distribution in the classification problem, where one tries to predict the type of some new object from the comparison of its features to those of a set of known objects.  The model is then applied to some well known examples of genomic inference from population statistics of an observable locus  We will conclude with a discussion of our findings and a summary of our results.

Some readers may find our use of the transformation group approach reactionary, archaic, or even naive, in light of the voluminous literature discussing other, more complicated strategies for deriving the form of the prior given some model for the likelihood of the observations, such as conjugate and entropic methods~\citep{raiffa-vn751926,lazo-2269298}.  Our response is that the analysis of similarity transformations has a long history in physics, leading one to statements of conservation of energy and momentum respective to translations in time and space.  When discussing the analysis of data, one should never forget that real measurements carry an index for location on the universal manifold and are subject to the laws of nature; how much use is made of that information depends upon the application and the investigator.

\section{Brief description of {B}ayesian data analysis}

The Bayesian approach to data analysis is best discussed using the language of conditional probability theory \citep{Bretthorst-1988,Durrett-1994,Sivia-1996}.  The expression for ``the probability of A given B'' can be written most compactly as \beq \label{eqn:pAgivenB}
p ( A \given B ) \equiv \p{A}{B} \; ,
\eeq where $A$ and $B$ can have arbitrary dimensionality; for example, $A$ could be a vector of measurements, and $B$ could include both the vector of parameters associated with some model as well as any other conditioning statements such as the model index.  The notation on the RHS of Equation~(\ref{eqn:pAgivenB}) is more economical than that of the LHS in terms of both the amount of ink on the page and the amount of mental effort required to keep track of the distinction between propositional statements in the superscript and conditional statements in the subscript; it also helps maintain identification of the units of density, which are carried by the propositional statements alone.  The sum and product rules of probability theory yield the expressions for marginalization and Bayes' theorem, \bea
\p{A}{} &=& \int_{\{B\}} dB \, \p{A , B}{} \;, \\
\p{B}{A} \p{A}{} &=& \p{A}{B} \p{B}{} \;,
\eea where marginalization follows from the requirement of unit normalization, and Bayes' theorem follows from requiring logical consistency of the joint density $\p{A , B}{} = \p{B , A}{}$.  Let us write as the vector $\mbf{m}$ the parameters for some model $M$, and let the data be written as $\mbf{x}$.  Bayes' theorem then relates the evidence for the parameters given the data $\p{\mbf{m}}{\mbf{x}}$ to the likelihood of the data given the parameters $\p{\mbf{x}}{\mbf{m}}$ through the expression \beq \label{eqn:bayesfordata}
\p{\mbf{m}}{\mbf{x}} \propto \p{\mbf{x}}{\mbf{m}} \p{\mbf{m}}{} \;,
\eeq where the factor $\p{\mbf{m}}{}$ describes the prior expectation over the parameter manifold in the absence of data, and the constant of proportionality $\p{\mbf{x}}{}$ represents the chance of measuring the data, which is usually recovered from the normalization requirement of the evidence density $\int_{\{\mbf{m}\}} d\mbf{m} \, \p{\mbf{m}}{\mbf{x}} = 1$.

The essential feature of Bayesian data analysis which takes it beyond maximum likelihood analysis is the inclusion of the prior density $\p{\mbf{m}}{}$.  The selection of the appropriate form of the prior for some coordinate mapping of the parameter manifold is guided by the principle of indifference applied to the behavior of the model under similarity transformations \citep{jaynes-1968,Sivia-1996,dose-350}.  Here, indifference is realized by examining the transformation group of the parameter manifold given by $\mbf{m}$.  Having found the prior measure for one coordinate system, the prior measure for alternate coordinate systems can be found through the use of a Jacobian transformation.

When only one model is in play, its quality of fit is irrelevant.  If no other description of the data is available, the most one can do is fit the parameters for the model at hand.  In order to accomplish the task of hypothesis testing, Bayesian data analysis forces one to specify explicitly the alternatives.  For a set of models indexed by $M$, the factors in Equation~(\ref{eqn:bayesfordata}) must be conditioned on the choice of $M$.  For two models $M \in \{ 1 , 2 \}$, the relative evidence is given by the ratio of the net evidence for each model, \beq \label{eqn:evdrat}
\rho^{1 \given \mbf{x}}_{2 \given \mbf{x}} \equiv \dfrac{\p{1}{\mbf{x}}}{\p{2}{\mbf{x}}} = \dfrac{\p{\mbf{x}}{1} \p{1}{}}{\p{\mbf{x}}{2} \p{2}{}} \; ,
\eeq where the factor $\p{1}{} / \p{2}{}$ describes any prior preference between the models and usually is identified as unity.  The factors in the likelihood ratio are given by the marginalization of the joint density over the parameter manifold for each model, \beq
\p{\mbf{x}}{M} = \int_{\{\mbf{m}\}} d\mbf{m} \, \p{\mbf{x} , \mbf{m}}{M} = \int_{\{\mbf{m}\}} d\mbf{m} \, \p{\mbf{x}}{\mbf{m} , M} \p{\mbf{m}}{M} \; ,
\eeq where the use of properly normalized densities for the likelihood and prior is required.  In particular, the prior $\p{\mbf{m}}{}$ is normalized to unity over the parameter manifold while the likelihood $\p{\mbf{x}}{\mbf{m}}$ retains its physical normalization.

An interesting feature of Bayesian model selection is that it accounts naturally for Occam's principle of efficiency.  Assuming model 1 has some parameter $a$ with uniform prior of extent $\Delta_a$, and taking the quadratic (Gaussian) approximation of its likelihood, without prior preference for either model the evidence ratio becomes \beq \label{eqn:Occam}
\dfrac{p(1 \given \mbf{x})}{p(2 \given \mbf{x})} = \dfrac{p(\mbf{x} \given a_0 , 1)}{p(\mbf{x} \given 2)} \left ( \dfrac{2 \pi \delta_a^2}{\Delta_a^2} \right )^{1/2} \; ,
\eeq where $a_0$ is the optimum value of the parameter and $\delta_a^2$ is its variance.  With an adjustable parameter, model 1 very likely provides a better quality of fit as measured by the first ratio $p(\mbf{x} \given a_0 , 1) / p(\mbf{x} \given 2)$; however, that is not the only factor in the net evidence ratio.  The improved fit to the data comes at the cost of the Occam factor $( 2 \pi \delta_a^2 / \Delta_a^2 )^{1/2}$ which measures the distribution of the evidence density relative to the parameter domain.  One requirement for the Gaussian approximation is that the prior not severely restrict the likelihood $\Delta_a \gg \delta_a$, thus the Occam factor works against the peak likelihood in the net evidence ratio in Equation~(\ref{eqn:Occam}).  Another interesting feature is that, all else being equal, the model whose parameters have the \textit{larger} variance is the one preferred by probability theory, as more of its parameter space is compatible with the measurements.  Suppose model 2 has its own parameter $b$ with comparable domain $\Delta_b \approx \Delta_a$ and provides a comparable fit to the data $p(\mbf{x} \given b_0 , 2) \approx p(\mbf{x} \given a_0 , 1)$.  In this case, the net evidence ratio reduces to $\p{1}{\mbf{x}} / \p{2}{\mbf{x}} \approx \delta_a / \delta_b$, so that the net evidence for model 1 relative to 2 is given by the ratio of the deviation of their parameters.

One criticism that is often leveled at those who use Bayesian methods~\citep{gelman-2008-33445} is that the ``prior and posterior [evidence] distributions represent subjective states of knowledge.''  By working in the language of conditional probability theory, what Bayesian methods require is that one specify the background knowledge upon which any inference of likelihood is based.  For example, one's estimate of the likelihood of rain today depends upon whether one has seen satellite images of clouds in the area.  Investigation of the transformation group associated with the parameters in a model leads one to specify the Haar measure as the intrinsic density which can serve as an objective prior in the absence of any further information.  The existence and uniqueness of the Haar measure hold under very general conditions on the set of parameters considered.

\section{Beta, Poisson, and gamma distributions in the continuum}
\label{sec:bpandg}

The beta distribution can be derived from consideration of the Poisson and gamma distributions in the continuum \citep{Press-1992,abramowitz-stegun}.  Physically, a continuum quantity is understood to be one for which the quantum unit is too small to measure.  Let us begin by supposing the amount $A$ for some quantity observed per unit time is given by a Poisson process with rate parameter $a$ expressed in the same physical units $u_a = u_A$, thus the likelihood can be written \beq
\p{A}{a} = a^A / e^a \Gamma (A+1) = a^A / e^a A \Gamma(A) \equiv \mathrm{Poisson} (A \given a) \; ,
\eeq in terms of the gamma function $\Gamma(A)$.  The discrete Poisson distribution is of course given by $\Gamma(A+1) \rightarrow A!$ for integer (quantized) $A$, such that the sum over all $A$ of the probability mass function is normalized, $e^{-a} \sum_{A = 0}^\infty a^A / A! = 1$.  One should keep in mind, however, that $\p{A}{a}$ is a probability density function which carries units of $u_A^{-1}$ such that $dA \, \p{A}{a}$ is a pure number.  The integral $\int_0^\infty dA \, \p{A}{a}$ cannot be easily evaluated; however, a collection of heuristic arguments (given in Appendix~\ref{sec:app}) indicate that its value also is unity.

According to \citet{jaynes-1968}, the parameter for a Poisson process must satisfy the same functional equation for transformations in scale as does the deviation parameter of a Gaussian distribution, thus the intrinsic (prior) density for $a \in [0, \infty]$ is given by \beq
\p{a}{} = a^{-1} / \int_0^\infty da \, a^{-1} \equiv a^{-1} / C_0 \; ,
\eeq which defines the infinite constant $C_0$.  Note that $C_0$ is formally equal to the mass of a distribution with infinite extent and unit density, $C_0 \equiv \int_\minfty^\infty dl$ for $l = \log a$, thus it also appears in the ubiquitous uniform prior of the maximum likelihood method.  Readers who are uncomfortable with infinite normalization constants may instead consider $C_0 \equiv \lim_{\epsilon \rightarrow 0} C_\epsilon$ for $C_\epsilon \equiv \int_{-\epsilon}^\epsilon dl$, using equivalent limits such that symmetry with respect to scale is maintained.  The intrinsic density $\p{a}{}$, whose sole proposition is the existence of $a$, is recognized as the Haar measure for the group of positive real numbers closed under the operation of multiplication.  Note that Jaynes' expression for the prior differs by a power from that obtained by application of the Jeffreys procedure, defined in terms of the square root of (the determinant of) the Fisher information (matrix).  That procedure yields the prior $\p{a}{F} \propto a^{-1/2}$ when applied to the Poisson distribution.  The Jaynes prior is functionally invariant under transformations of the form $\alpha = m a^n$ for given $m$ and $n$, such that $\p{\alpha}{} \propto \alpha^{-1}$, whereas $\p{\alpha}{F} \propto \alpha^{(1-2n)/2n}$ which is invariant only for $n = 1$.  In the limit $n \rightarrow \infty$ one finds $\p{\alpha}{F} \rightarrow \p{\alpha}{}$, which can be interpreted heuristically (but maybe not correctly) as follows.  When evaluating the Fisher information, the expectation value is taken over only a single datum, whereas the measurement process could be repeated any number of times, which for the Poisson process amounts to changing the unit of time.  It seems, then, that the Jaynes prior accounts for the possibility of an infinite number of measurements when assigning the most general form of $\p{a}{}$.

The joint density over the manifold $(a,A)$ can be written as the product of the conditional density $\p{A}{a}$ and intrinsic density $\p{a}{}$, \beq
\p{a,A}{} = \p{A}{a} \p{a}{} = a^{A-1} / C_0 e^a A \Gamma(A) \; ,
\eeq and its integral over $a$ can be evaluated explicitly, \beq
\int_0^\infty da \, \p{a,A}{} = \int_0^\infty da \, \p{A}{a} \p{a}{} = A^{-1} / C_0 \equiv \p{A}{} \; ,
\eeq which is recognized as the chance of measuring $A$.  Having equivalent physical units, the quantities $a$ and $A$ possess the same transformation group, thus their intrinsic densities must be functionally identical.  That the expression $\p{a,A}{}$ represents a valid probability density function is verified by next taking the integral over $A$, \beq
\int_0^\infty dA \int_0^\infty da \, \p{a,A}{} = \int_0^\infty dA \, \p{A}{} = 1 \; ,
\eeq thus the joint density has unit mass over the infinite quarter plane $[0,\infty] \times [0,\infty]$ in $\mathbb{R}^2$.

According to Bayes' theorem, the evidence for parameter $a$ conditioned on observable $A$ in the Poisson likelihood is given by the gamma distribution, \beq
\p{a}{A} = \p{A}{a} \p{a}{} / \p{A}{} = a^{A-1} / e^a \Gamma(A) \equiv \mathrm{Gamma} (a \given A) \; ,
\eeq which is normalized to unit mass, $\int_0^\infty da \, \p{a}{A} = 1$.  One also can verify the integral $\int_0^\infty da \, \p{A}{a} = 1$ (permissible since $u_a = u_A$), thus the likelihood is normalized over the parameter $a$ as well; logically, given the existence of a value for $A$, it must be true that the sum of all its conditional probabilities is equal to unity.  By similar logic, the normalization of both $\p{A}{a}$ and $\p{a}{A}$ over $A$ should also be true, but a direct evaluation of those integrals analytically is difficult (see Appendix~\ref{sec:app}).  For comparison, consider the joint density of a measurement $M$ and parameter $m$ given by a Gaussian of known deviation which sets the scale, $\p{m,M}{} = \exp [-\pi (M-m)^2] / C_0$, with intrinsic densities $\p{m}{} = \p{M}{} = 1 / C_0$.  In this case, one can easily show that $\int_\minfty^\infty dm \, \p{m}{M} = \int_\minfty^\infty dm \, \p{M}{m} = 1$ as well as $\int_\minfty^\infty dM \, \p{M}{m} = \int_\minfty^\infty dM \, \p{m}{M} = 1$.  That similar normalizations hold in the continuum for the Poisson and gamma densities is the main conjecture of this paper.

Note that the joint density $\p{a,A}{}$ does not care whether $a$ and $A$ are identified as parameter and observable, respectively, or \textit{vice versa}.  The identification of evidence, chance, likelihood, and prior similarly is arbitrary, as long as one is consistent \citep{Sivia-1996}.  The decomposition through Bayes' theorem of the joint density in terms of intrinsic densities given by the Haar measure allows one to write \beq \label{eqn:aPAG}
a^{-1} \mathrm{Poisson} (A \given a) = A^{-1} \mathrm{Gamma} (a \given A) \; ,
\eeq thus the gamma distribution is the evidence for a Poisson process likelihood, and \textit{vice versa}.  The second (shape) parameter commonly associated with the gamma distribution can be identified as the ratio of the units for the parameter and observable $u_a / u_A$, which here is specified as unity.

Now let us consider the joint density $\p{a,b,A,B}{}$, which can be written as \beq
\p{A,B}{a,b} \p{a,b}{} = \left[ a^A b^B / e^{a+b} A B \Gamma(A) \Gamma(B) \right] / a b C_0^2 \; .
\eeq  Under a change of coordinate mapping $(a,b) \rightarrow (x,y)$ such that \beq
\left[ \begin{array}{c} x \\ y \end{array} \right] = \left[ \begin{array}{c} a / (a+b) \\ a+b \end{array} \right] \Longleftrightarrow \left[ \begin{array}{c} a \\ b \end{array} \right] = \left[ \begin{array}{c} x y \\ (1-x) y \end{array} \right] \; ,
\eeq with domain $x \in [0, 1]$ and $y \in [0, \infty]$, the Jacobian matrix is given by \beq
\msf{J}^{a,b}_{x,y} \equiv \dfrac{\partial (a,b)}{\partial (x,y)} = \left[ \begin{array}{cc} y & x \\ -y & 1-x \end{array} \right] \; ,
\eeq whose determinant is $\abs{\msf{J}^{a,b}_{x,y}} = y$.  The intrinsic density in the new coordinates is thus \beq
\p{x,y}{} = \p{a,b}{} \abs{\msf{J}^{a,b}_{x,y}} = x^{-1} (1-x)^{-1} y^{-1} / C_0^2 \; ,
\eeq and the conditional density is \beq
\p{A,B}{x,y} = x^A (1-x)^B y^{A+B} / e^y A B \Gamma(A) \Gamma(B) \; .
\eeq  Since $\p{A,B}{} = 1 / A B C_0^2$, one can then write \beq
\p{x,y}{A,B} = \p{A,B}{x,y} \p{x,y}{} / \p{A,B}{} = x^{A-1} (1-x)^{B-1} y^{A+B-1} e^{-y} / \Gamma(A) \Gamma(B) \; ,
\eeq which integrates to unity, \beq
\int_0^1 dx \int_0^\infty dy \, \p{x,y}{A,B} = 1 \; ,
\eeq using the evaluations \bea
\int_0^\infty dy \, y^{A+B-1} e^{-y} &=& \Gamma(A+B) \; , \\
\int_0^1 dx \, x^{A-1} (1-x)^{B-1} &=& \beta(A,B) \; .
\eea  Marginalization then yields \bes
\p{x,A,B}{} &=& \int_0^\infty dy \, \p{x,y,A,B}{} = x^{A-1} (1-x)^{B-1} / \beta(A,B) A B C_0^2 \\
 &=& \p{x}{A,B} \p{A,B}{} \; ,
\ees which is the main result of this section.  With the interpretation of $x = a / (a+b) \in [0, 1]$ as a normalized frequency (rate of observance), one can state that the intrinsic density for an absolute likelihood is $\p{x}{} = x^{-1} (1-x)^{-1} / C_0$, while that for a relative likelihood $r = a/b \in [0, \infty]$ is $\p{r}{} = r^{-1} / C_0$.  Note that $C_0$ is infinite only when the parameter is allowed to obtain the extreme values of its domain, and in fact is comprised of two independent infinities $C_0 = 2 \int_1^\infty dr \, r^{-1}$, one from each boundary of the manifold.

While the relationship between these three distributions has been explored by many authors, nowhere have we found a derivation within the framework of conditional probability theory that ties them together under the conjecture of the continuum normalization.  The literature has instead focused on the relation between discrete random variables rather than the continuous case.  Partly that may be because the expression of the Poisson distribution in the continuum is not so widely known, owing to the difficulty of evaluating its normalization integral analytically.  Another reason may be because use of transformation group arguments has been championed primarily by physicists rather than statisticians.  Whatever the reason, the establishment of Equation~(\ref{eqn:aPAG}) in the continuum leads one naturally to the beta distribution, which displays explicitly the transformation group prior for the normalized frequency $x$.

Let us now consider the parametrization $(x,y) \rightarrow (\alpha,\beta)$ given by\beq
\left[ \begin{array}{c} \alpha \\ \beta \end{array} \right] = \left[ \begin{array}{c} \log y \\ \log x - \log (1-x) \end{array} \right] \Longleftrightarrow \left[ \begin{array}{c} x \\ y \end{array} \right] = \left[ \begin{array}{c} 1/[1 + e^{-\beta}] \\ e^\alpha \end{array} \right] \; ,
\eeq with domain $\alpha, \beta \in [-\infty, \infty]$.  In these coordinates, the prior density is uniform $\p{\alpha,\beta}{} = 1 / C_0^2$, thus the evidence is proportional to the likelihood, and the joint density equals \bes
\p{A,B}{\alpha,\beta} \p{\alpha,\beta}{} &=& \left[ x^A (1-x)^B y^{A+B}  / e^y A B \Gamma(A) \Gamma(B) \right] / x (1-x) y C_0^2 \\
 &=& \left[ 1 + e^{-\beta} \right]^{-A} \left[ 1 + e^\beta \right]^{-B} e^{(A+B) \alpha} \left\lbrace \exp (e^\alpha) A B \Gamma(A) \Gamma(B) \right\rbrace^{-1} / C_0^2 \\
  &=& \left\lbrace \dfrac{e^\alpha}{1+e^{-\beta}} \right\rbrace^A \left\lbrace \dfrac{e^\alpha}{1+e^\beta} \right\rbrace^B \left\lbrace \exp (e^\alpha) A B \Gamma(A) \Gamma(B) \right\rbrace^{-1} / C_0^2 \\
  &=& \left[ e^{-\alpha} + e^{-\alpha-\beta} \right]^{-A} \left[ e^{-\alpha} + e^{-\alpha+\beta} \right]^{-B} \left\lbrace \exp (e^\alpha) A B \Gamma(A) \Gamma(B) \right\rbrace^{-1} / C_0^2 \; .
\ees  The first two factors above are reminiscent of the logistic regression model \citep{pengandso-0101_04}; however, the parameter $\alpha$, commonly called ``the intercept'', makes an appearance as the argument of a double exponential in the third factor as well as in the terms $e^{-\alpha}$ without $\beta$.  The third factor is not related to the prior thus must be part of the likelihood.  Rather than conflating the parameters, keeping the likelihood models $\p{A,B}{\alpha} \propto e^{(A+B) \alpha} / \exp e^\alpha$ and $\p{A,B}{\beta} \propto [ 1 + e^{-\beta} ]^{-A} [ 1 + e^\beta ]^{-B}$ independent leads to a more efficient evaluation \citep{johnson-ws2017}.

\section{Application to prediction and classification}

Let us begin this section by talking about baseball.  Specifically, let us consider the use of the seasonal batting average as a predictor for whether a player will reach base on his next appearance.  Let each appearance be indexed by time given by integer $t \in [1,T]$, and let us identify a successful appearance as an event of type $A$, while outs are of type $B$.  The record of successful appearances can be notated by $\mbf{A} \equiv A_j$ for $j \in [1,J]$, and similarly for $\mbf{B} \equiv B_k$ of dimension $K$, such that $T = J + K$.  The evidence for the value of the batting average $x$ is the product of the prior and likelihood factors, yielding the beta distribution $\p{x}{J,K} \propto x^{J-1} (1-x)^{K-1}$ with mode $x_E = (J-1) / (J+K-2)$ and expectation value $\mean{x}{x \given J,K} = J / (J+K)$, which coincides with the likelihood mode $x_L$ and gives the predicted rate of success for the next appearance.

One can incorporate into the form of the prior $\p{x}{}$ additional information pertinent to the problem at hand.  In particular, one can use knowledge of the seasonal nature of the sport to impose sensible limits on the domain $x \in [\epsilon,1-\epsilon]$.  If our player's season is not yet over, then there must be at least one more at bat scheduled.  A sensible limit is thus given by $\epsilon = 1/(T+1)$, which incorporates the notions that nobody is perfect (1 is excluded) and of the benefit of the doubt (0 is excluded); assuming our player is a professional at least one event of each type should be observed per season, even for pitchers.  One effect of such a prior is that it does not allow observations of only one type of event to pull the evidence mode all the way to the hypothetical limits of 0 and 1.  Another effect is that early in the season $T \gtrsim 1$ the domain of $x$ requires an observation of the batter before starting to make predictions; once we are certain the batter is playing this season $T = 1$, we can state the expected chance of success is equal to 1/2, the only allowed point, with further observations expanding the domain until at the end of a long season $T \gg 1$ the prior is wide open.

Let us now turn to consideration of classifying some new event as type $A$ or $B$ on the basis of its location relative to those for $T$ observations whose classification is assigned.  The elements of the measurement vectors $\mbf{A}$ and $\mbf{B}$ are now locations along some axis $\tau$, with a measurement uncertainty expressed by the Gaussian deviation $\sigma$.  If the chance an event is of type $A$ is independent of location, one can write $\p{x,\tau}{\sigma,\mbf{A},\mbf{B}} \propto \p{x}{J,K} \p{\tau}{\sigma,\mbf{A},\mbf{B}}$, where $\p{\tau}{\sigma,\mbf{A},\mbf{B}}$ is a Gaussian centered on the mean location of all the events and each margin is normalized independently.  That is obviously not the solution we are looking for, which should give an expectation of the form $x(\tau)$ based on a joint density that can be factored as $\p{x,\tau}{\sigma,\mbf{A},\mbf{B}} = \p{x}{\sigma,\tau,\mbf{A},\mbf{B}} \p{\tau}{}$ for $\p{\tau}{} \propto 1$.

Another way to express the notion that location has become irrelevant is by taking the limit $\sigma \rightarrow \infty$.  In that case, one should require $\p{x}{\sigma,\tau,\mbf{A},\mbf{B}} \rightarrow \p{x}{J,K}$ for all $\tau$, which corresponds to neglecting the stadium of appearance in the batting average problem above.  In doing so, we have not said that location does not exist, but rather that location does not matter.  For finite $\sigma$, we should write $\p{x}{\sigma,\tau,\mbf{A},\mbf{B}} \propto \p{x}{} \p{\mbf{A},\mbf{B}}{\sigma,\tau,x}$, whose limit for $\tau \rightarrow \infty$ is $\p{x}{}$; observations nearby should not significantly affect our prediction for a galaxy far, far away.  The problem now is one of assigning the appropriate form for the likelihood factor.  For inspiration, we have looked at various approaches suggested in the literature \citep{terrell-1236,hall-5276H,kim-2529,eberts-2013}.

\begin{figure}[]
\begin{center}
\includegraphics[width=.8\textwidth]{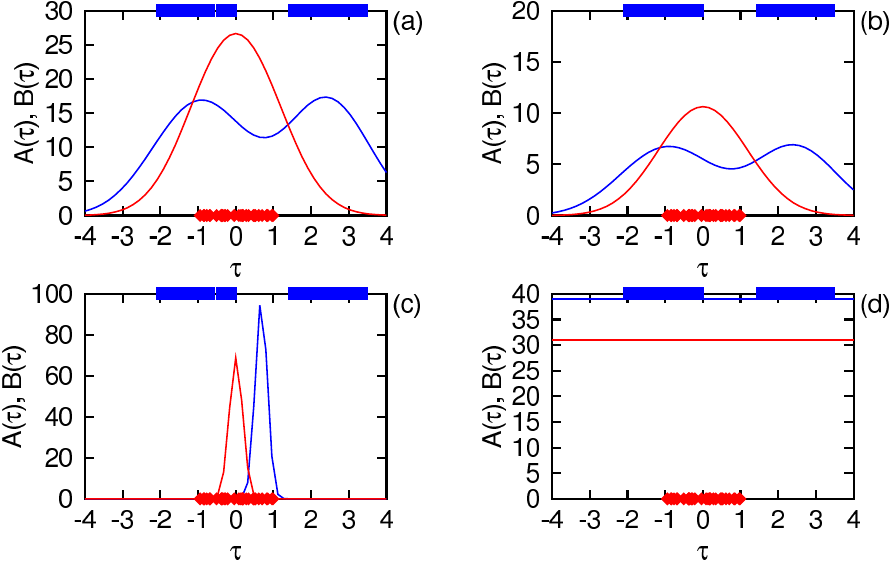}
\end{center}
\caption{Distributions $A(\tau)$ and $B(\tau)$ as described in the text.  The locations $A_j$ are indicated at the top of each plot, and $B_k$ are at the bottom.}
\label{fig:B}
\end{figure}

At this stage the discussion becomes a bit heuristic.  When the observations are independent, we can factor the likelihood into the form \beq
\p{\mbf{A},\mbf{B}}{\sigma,\tau,x} = \prod_j \p{j}{\sigma,\tau,x} \prod_k \p{k}{\sigma,\tau,x} \; ,
\eeq where $\p{j}{\sigma,\tau,x}$ represents the chance datum $j$ is of type $A$, and similarly for $\p{k}{\sigma,\tau,x}$.  What, then, is the form of $\p{j}{\sigma,\tau,x}$ that yields sensible results for all $\sigma$ and irrespective of the underlying spatial distributions of the two types of events?  A form which suggests itself is more clearly notated in terms of its logarithm $\q{j}{\sigma,\tau,x} = - r^j_\tau \log x$, where $r^j_\tau = \exp^{-1/2} [(A_j - \tau)^2 / \sigma^2]$ is the probability of an event at $A_j$ relative to that at $\tau$.  The log of the likelihood can then be written as \beq \label{eqn:qAB_stx}
-\q{\mbf{A},\mbf{B}}{\sigma,\tau,x} = \sum_j r^j_\tau \log x + \sum_k r^k_\tau \log (1-x) \; ,
\eeq whose limits are $J \log x + K \log (1-x)$ for $\sigma \rightarrow \infty$ and 0 for $\tau \rightarrow \infty$, in accord with our requirements for the evidence density.  Let us identify $A(\tau) \equiv \sum_j r^j_\tau$, and similarly for $B(\tau)$; then the likelihood can be written as $x^{A(\tau)} (1-x)^{B(\tau)}$, and the evidence for the value $x$ at $\tau$ is given by \beq
\p{x}{\sigma,\tau,\mbf{A},\mbf{B}} \propto x^{A(\tau)-1} (1-x)^{B(\tau) - 1} \; ,
\eeq which has the form of a beta distribution at all locations.  An example of $A(\tau)$ and $B(\tau)$ for an arbitrary distribution of $\mbf{A}$ and $\mbf{B}$ in units of the deviation $\sigma = 1$ is shown in panel (a) of Figure~\ref{fig:B}.  The values $A_j$ are drawn uniformly over two disjoint regions each with a span of 2 units, and the values $B_k$ are selected from a region spanning 2 units which overlaps partially one of the type $A$ regions.

Out of respect for our heuristic argument, we should consider some alternative definitions for the likelihood.  If instead of the relative probabilities $r^j_\tau$ one defines $A(\tau)$ as the sum of the absolute probabilities $\p{j}{\sigma,\tau} = (2 \pi \sigma^2)^{-1/2} r^j_\tau$ such that $\int d\tau \sum_j \p{j}{\sigma,\tau} = J$, one has in the limit $\sigma \rightarrow \infty$ the result $A(\tau) \rightarrow 0$, which does not recover the beta distribution in terms of $J$ and $K$.  If one uses the product of the datum likelihoods to define $A(\tau) = J (2 \pi \sigma^2 / J)^{-1/2} \exp^{-1/2} [(\tau - \mu_A)^2 J / \sigma^2]$ for $\mu_A = \mean{A_j}{j}$, which also integrates over $\tau$ to $J$, one's estimate for the evidence depends upon only the first moments of the event distributions, a procedure which is easily foiled when the underlying location distribution are not Gaussian.  Finally, if one uses $\p{j}{\sigma,\tau,x} = x \p{j}{\sigma,\tau}$, one recovers simply the independent distributions over $x$ and $\tau$.  Examples of these definitions of $A(\tau)$ and $B(\tau)$ are displayed in panels (b) through (d) respectively of Figure~\ref{fig:B} for the same distributions of $A_j$ and $B_k$.

\begin{figure}[]
\begin{center}
\includegraphics[width=.8\textwidth]{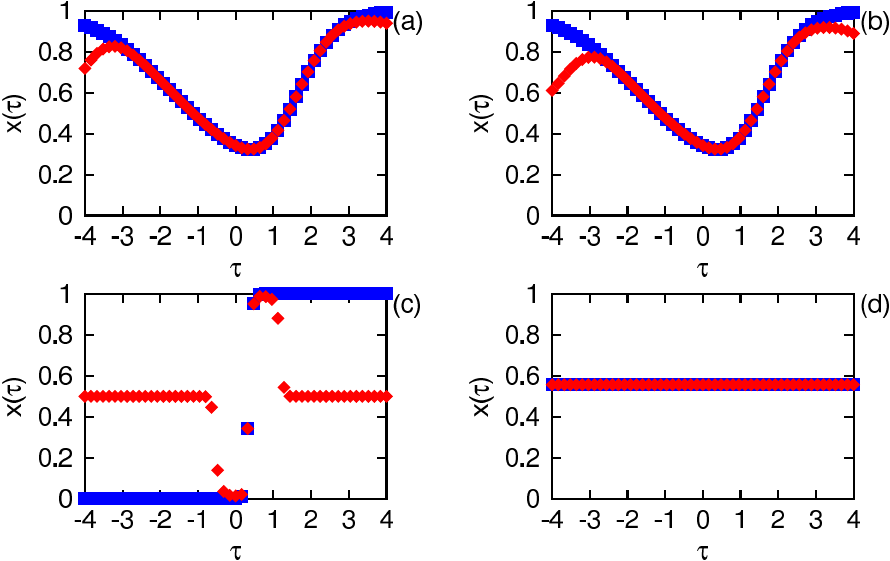}
\end{center}
\caption{Prediction values $x(\tau)$ from $A(\tau)$ and $B(\tau)$ as described in the text.  The maximum likelihood predictor $x_\mathrm{ML}$ is shown as $\square$, and the expectation value $x_\mathrm{EV}$ is shown as $\lozenge$.}
\label{fig:C}
\end{figure}

A maximum likelihood predictor can be formed from the expression \beq
x_\mathrm{ML}(\tau) = [1 + B(\tau) / A(\tau)]^{-1} \; ,
\eeq which is evaluated from the measurements $\mbf{A}$ and $\mbf{B}$ with respect to $\sigma$.  The expectation value $x_\mathrm{EV}(\tau) = \mean{x}{x \given \sigma,\tau,\mbf{A},\mbf{B}}$, however, takes into account the full domain of $x$ as measured by the evidence density.  In Figure~\ref{fig:C} we display the maximum likelihood and expected value predictors for the distributions $A(\tau)$ and $B(\tau)$ shown in Figure~\ref{fig:B}.  The likelihood estimate $x_\mathrm{ML}$ is the same in panels (a) and (b), since the ratio $B(\tau) / A(\tau)$ in terms of the summed likelihoods does not depend on their normalization.  The expectation value in panel (b) is more conservative, in that it more quickly approaches the expectation value of the prior, compared to panel (a).  The likelihood estimate for the method of panel (c) gives a prediction for the region $\tau \in [-2,-1]$ that is contrary to the observations, while its expectation value is very quickly drawn to that of the prior, even in the region $\tau > 1.5$ where only type $A$ events are observed.  The likelihood and expectation value predictors are identical when the location information is ignored, as seen in panel (d).

\begin{figure}[]
\begin{center}
\includegraphics[width=.8\textwidth]{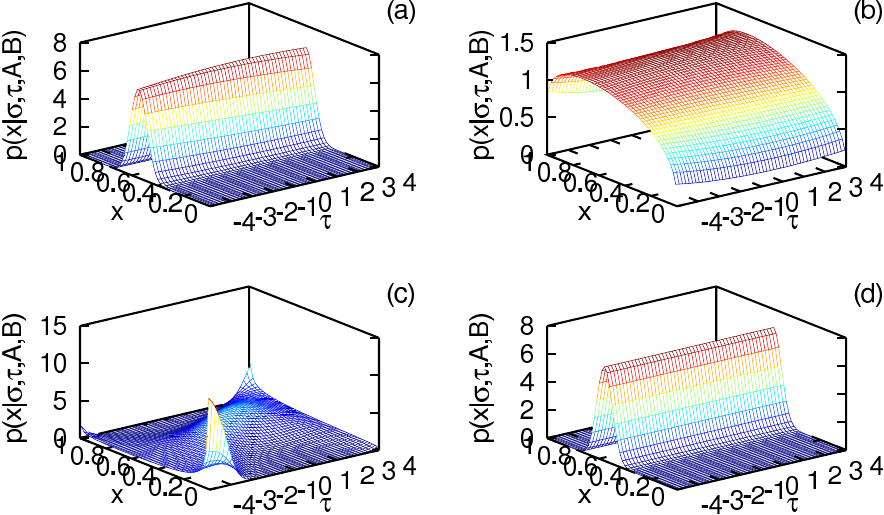}
\end{center}
\caption{Evidence densities $\p{x}{\sigma,\tau,\mbf{A},\mbf{B}}$ for $\sigma = 10$ as described in the text.}
\label{fig:D}
\end{figure}

Let us now repeat the evaluation of the evidence densities $\p{x}{\sigma,\tau,\mbf{A},\mbf{B}}$ for the various definitions of $A(\tau)$ and $B(\tau)$, but this time let us suppose that $\sigma = 10$ for the same locations $\mbf{A}$ and $\mbf{B}$.  Let us also inspect the evidence densities directly, to see which one best encodes a reasonable estimate of the solution to our problem.  In Figure~\ref{fig:D} we display the evidence density for $x$ as a function of $\tau$ for the various likelihood models.  We can see that panel (a) is the one most like panel (d), which evaluates the beta distribution without regard to location.  The other models, panels (b) and (c), are not in accord with the conclusions a reasonable observer would draw intuitively from the presented data; surely with close to 100 observations the relative rate of production should be fairly well determined over the common region of the events.  While our justification of Equation~(\ref{eqn:qAB_stx}) is heuristic, its form is the same as that of a Gaussian with unequal weights, where each datum factor in the likelihood is an absolute probability to the power of a relative probability.

What can we say about the limit $\sigma \rightarrow 0$, which indicates that observations are relevant only to predictions at the same location?  With respect to the finite resolution of whatever apparatus is used to take the location measurements, what we really mean in that limit is that locations are resolved over a set of discrete channels which have no influence or bearing on events in other channels.  Returning to the baseball analogy, that model asserts that batting averages for each stadium should be evaluated independently, which is not an unreasonable procedure, given by $A(\tau) = \sum_{A(j) = \tau} r^j_\tau$ and similarly for $B(\tau)$.  The parameter $\tau$ can in fact be an abstract location, not just a physical one, with the interpretation of $r^j_\tau$ as the relevance of observations in one channel to predictions in another.  We should also point out that we have been treating the location $\tau$ of the predicted classification as a quantity known exactly; if the location of the unclassified event $\tau\prime$ is itself subject to measurement deviation $\sigma$, then one must convolute the evidence density with its normalized distribution, $\p{x}{\sigma,\tau\prime,\mbf{A},\mbf{B}} = \int d\tau \p{\tau}{\sigma,\tau\prime} \p{x}{\sigma,\tau,\mbf{A},\mbf{B}}$.  Furthermore, if the value of $\sigma$ is unknown, it can be integrated out by treating it as as a nuisance parameter, $\p{x}{\tau\prime,\mbf{A},\mbf{B}} = \int d\sigma \p{\sigma}{} \p{x}{\sigma,\tau\prime,\mbf{A},\mbf{B}}$ for $\p{\sigma}{} \propto \sigma^{-1}$.

\section{Application to the {Balding-Nichols} model}

Next let us look at how the beta distribution is used in the analysis of genetic profiles.  Suppose the gene at some locus has a dominant allele $G$ and a recessive allele $g$ such that the genotypes $GG$, $Gg$, and $gg$ are distinguishable.  According to \citet{balding-963}, the allele frequency $x$ for finding $G$ at the locus follows a beta distribution with parameters $A = \mu (1-\lambda) / \lambda$ and $B = (1-\mu) (1-\lambda) / \lambda$.  The measurements are now not values for $x_k$ but rather the number of members of each genotype observed within a sampling of the $k$th population, $N_k = N_{k,GG} + N_{k,Gg} + N_{k,gg}$.  In terms of the parameters, the probability for an individual to be a member of the genotype is given by \bes
\p{GG}{A,B} \equiv \langle x^2 \rangle_{x \given A,B} &=& A (A+1) / (A+B) (A+B+1) \\
 &=& \lambda \mu + (1-\lambda) \mu^2
\ees for the dominant homozygote, and by \bes 
\p{gg}{A,B} \equiv \langle (1-x)^2 \rangle_{x \given A,B} &=& B (B+1) / (A+B) (A+B+1) \\
 &=& \lambda (1-\mu) + (1-\lambda) (1-\mu)^2
\ees for the recessive homozygote, while the heterozygote appears with probability \bes
\p{Gg}{A,B} \equiv 2 \langle x (1-x) \rangle_{x \given A,B} &=& 2 A B / (A+B) (A+B+1) \\
 &=& 2 (1-\lambda) \mu (1-\mu) \; ,
\ees where the factor of 2 accounts for the indistinguishability of the order of the alleles.  In matrix form with unit 1-norm, the joint distribution of the genotypes can be written \beq
\left[ \begin{array}{cc} \p{GG}{\lambda,\mu} & \p{Gg}{\lambda,\mu} / 2 \\ \p{Gg}{\lambda,\mu} / 2 & \p{gg}{\lambda,\mu} \end{array} \right] = \lambda \left[ \begin{array}{cc} \mu & 0 \\ 0 & 1-\mu \end{array} \right] + (1-\lambda) \left[ \begin{array}{cc} \mu^2 & \mu (1-\mu) \\ \mu (1-\mu) & (1-\mu)^2 \end{array} \right] \; ,
\eeq yielding the interpretation of $\mu = (1+B/A)^{-1}$ as the mean dominant allele frequency and of $\lambda = (1+A+B)^{-1}$ as a measure of heterozygote suppression.  The parameter $\lambda$ may be identified with Wright's inbreeding coefficient $F$.

The probability of obtaining the measurements given knowledge of the parameter values is the product of the genotype likelihoods weighted by the number of members.  For a single population, \beq
\p{N_{GG},N_{Gg},N_{gg}}{A,B} = (\p{GG}{A,B})^{N_{GG}} (\p{Gg}{A,B})^{N_{Gg}} (\p{gg}{A,B})^{N_{gg}} \; ,
\eeq thus the information content of the data (negative log likelihood) is \beq \label{eqn:Llm}
L(A,B) \equiv \q{N_{GG},N_{Gg},N_{gg}}{A,B} = N_{GG}\, \q{GG}{A,B} + N_{Gg}\, \q{Gg}{A,B} + N_{gg}\, \q{gg}{A,B} \; ,
\eeq recalling $q \equiv - \log p$.  The nontrivial solution of $\del L(A,B) = 0$ yields the maximum likelihood estimate of the optimal parameter values \beq
\left[ \begin{array}{c} a_L \\ b_L \end{array} \right] = \left[ \begin{array}{c} (2 N_{Gg} N_{GG} + N_{Gg}^2) / (4 N_{GG} N_{gg} - N_{Gg}^2) \\ (2 N_{Gg} N_{gg} + N_{Gg}^2) / (4 N_{GG} N_{gg} - N_{Gg}^2) \end{array} \right] \; ,
\eeq which corresponds to the location \beq
\left[ \begin{array}{c} \lambda_L \\ \mu_L \end{array} \right] = \left[ \begin{array}{c} (4 N_{GG} N_{gg} - N_{Gg}^2) / [ (2 N_{gg} + N_{Gg}) (2 N_{GG} + N_{Gg}) ] \\ (2 N_{GG} + N_{Gg}) / 2 (N_{GG} + N_{Gg} + N_{gg}) \end{array} \right]
\eeq on the $(\lambda,\mu)$ manifold.  Suppose now instead of the genotype observations our data consists of the raw allele counts for $G$ and $g$, given by $N_G \equiv 2 N_{GG} + N_{Gg}$ and $N_g \equiv 2 N_{gg} + N_{Gg}$ such that $2 N = N_G + N_g$.  The log likelihood in this case becomes \bes
\q{N_G,N_g}{A,B} &=& N_G \log (1+B/A) + N_g \log (1+A/B) \\
 &=& - N_G \log \mu - N_g \log (1-\mu) \; ,
\ees whose optimal estimate is the same $\mu_L = (1+N_g/N_G)^{-1}$ with $\lambda$ undetermined.  From the raw allele counts one can resolve only the dominant allele frequency for a single population.

\begin{figure}[]
\begin{center}
\includegraphics[scale=0.75]{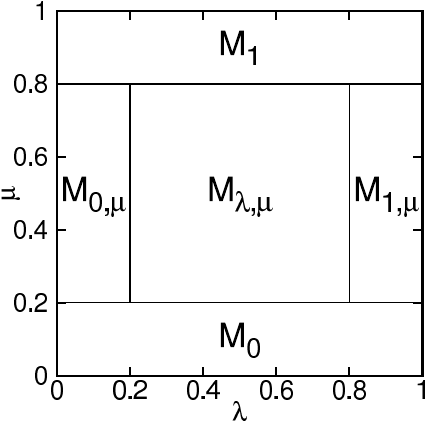}
\end{center}
\caption{Mapping from the parameter manifold $(\lambda,\mu)$ to the model labels described in the text for an exaggerated value of $\epsilon$.}
\label{fig:A}
\end{figure}

The merit function for the evidence density $\p{\lambda,\mu}{N_{GG},N_{Gg},N_{gg}}$ in terms of the parameters $(\lambda,\mu)$ can be written as \beq
F(\lambda,\mu) = L(\lambda,\mu) + \log [\lambda (1-\lambda) \mu (1-\mu)] \; ,
\eeq using an unnormalized prior.  When $\lambda = 0$, the population is said to be in Hardy--Weinberg equilibrium with a single parameter $\mu$ for the dominant allele frequency; however, one should observe that $\lambda = 1$ is also an equilibrium solution with a single parameter $\mu$.  Those two cases correspond to the peaks in the prior for $\lambda$ when the boundary is not excluded.  In the limit $\epsilon \rightarrow 0$, the normalized prior $\p{\lambda}{}$ has the value 1/2 at $\lambda$ equal to 0 or 1 and the value 0 everywhere else.  Similarly, when $\mu$ equals 0 or 1, one finds that $\lambda$ is undetermined by the likelihood, thus those models have zero free parameters.  The five models under consideration (for a single population) can thus be labeled $M_{\lambda,\mu}$, $M_{0,\mu}$, $M_{1,\mu}$, $M_0$, and $M_1$, where the first is a two parameter model, the next two are one parameter models, and the last two zero parameter models, all of which are conditioned on the value of the boundary exclusion $\epsilon$ determined in principle by the nature of the measurement apparatus.  A similar approach is suggested by \citet{johros-RSSB730}.  See Figure~\ref{fig:A} for a depiction of the mapping from the parameter manifold to the model labels using a large value of $\epsilon$ for clarity.

It is instructive to look at the information content of the data with respect to the various models.  For $M_{\lambda,\mu}$ with two parameters, $L_{\lambda,\mu}$ is given by Equation~(\ref{eqn:Llm}), whose mode provides a good starting point for the numerical optimization of $F_{\lambda,\mu}$; four other points to consider are the projections of the likelihood mode onto the boundaries of the manifold.  The model $M_{0,\mu}$ has an information density of \beq
L_{0,\mu} (\mu) = - N_{Gg} \log 2 - (2 N_{GG} + N_{Gg}) \log \mu - (2 N_{gg} + N_{Gg}) \log (1-\mu) \; ,
\eeq retaining the constant term with $N_{Gg}$, and $M_{1,\mu}$ has \beq
L_{1,\mu} (\mu) = - N_{Gg} \log 0 - N_{GG} \log \mu - N_{gg} \log (1-\mu) \; ,
\eeq supported only when $N_{Gg} = 0$ such that $N_{Gg} \log \p{Gg}{\lambda = 1} = \log 0^0 = 0$; otherwise, $L_{1,\mu} = \infty$.  For either one parameter model, it is possible for certain values of the input data to yield an evidence density which is uniform in $\mu$; in those cases, the mode is undetermined and the unnormalized evidence density is equal to 1.  For the zero parameter models, \beq
L_0 = - N_{GG} \log 0 - N_{Gg} \log 0 - N_{gg} \log 1 \; ,
\eeq which equals 0 when only $N_{gg} > 0$ else is infinite, and by symmetry \beq
L_1 = - N_{GG} \log 1 - N_{Gg} \log 0 - N_{gg} \log 0 \; .
\eeq  Since the zero parameter models have a manifold of a single point, their net evidence (mean likelihood) is either 0 or 1 according to whether they are supported by the data, which sets the unit of evidence when comparing the other models.

\begin{table}
\caption{Genotype observations from \citet{ford-1971} and maximum likelihood results with values for $\lambda$, $\mu$, and $P$ stated in units of percent}
\label{tab:D} 
\centering
\begin{tabular}{*{8}{c}}
\hline
$N_{GG}$ & $N_{Gg}$ & $N_{gg}$ & $\lambda_L$ & $\mu_L$ & $\chi^2_P$ & $P_1(\chi^2_P)$ \\
\hline
1469 & 138 & 5 & 2.270 & 95.409 &  0.831 &  63.8 \\
\hline
\end{tabular}
\end{table}

\begin{table}
\caption{Evidence analysis of the data from Table~\ref{tab:D} with values for $\lambda$, $\mu$, and $\bigq{}{}$ stated in units of percent}
\label{tab:E} 
\centering
\begin{tabular}{l *{5}{c}}
\hline
model: & $\lambda,\mu$ & $0,\mu$ & $1,\mu$ & $0$ & $1$ \\      
\hline                                                         
mode: & (0.001,95.438) & 95.438 &     NaN & NaN & NaN \\       
mean: & (0.985,95.407) & 95.409 &     NaN & NaN & NaN \\       
$\q{M}{\mbf{N}}$: & 510.4 & 509.6 &   Inf &   Inf &   Inf \\   
$\bigq{M}{\mbf{N}}$: &  30.2 &  69.8 &   0.0 &   0.0 &   0.0 \\
\hline
\end{tabular}
\end{table}

As an illustration, let us look first at some data from \citet{ford-1971} shown in Table~\ref{tab:D}.  Also shown are the maximum likelihood values $\lambda_L$ and $\mu_L$ in units of percent.  From these numbers one can evaluate Pearson's statistic $\chi^2_P$ from the Hardy--Weinberg expectation values $N_{GG}^{HW} = N \mu_L^2$ and so on.  The accumulation of the $\chi^2_P$ statistic for 1 degree of freedom (3 from the data less 2 used in the model) gives the significance $P_1(\chi^2_P)$ of the deviation from equilibrium, and for comparison $P_1(3.84) \approx 95\%$ for $P_d(\chi^2) \equiv \gamma(d/2, \chi^2/2) / \Gamma(d/2)$ in the notation used by \citet{Press-1992}.  The conventional interpretation is to state that the equilibrium model is not rejected on account of the small value of $\chi^2_P$; however, since only two models are considered, one may interpret the value of $P_1(\chi^2_P)$ as the amount of probability not assigned to the equilibrium model, in which case the maximum likelihood analysis is showing   some preference, if not overwhelming, for the non-equilibrium model.

\begin{figure}[]
\begin{center}
\includegraphics[scale=1]{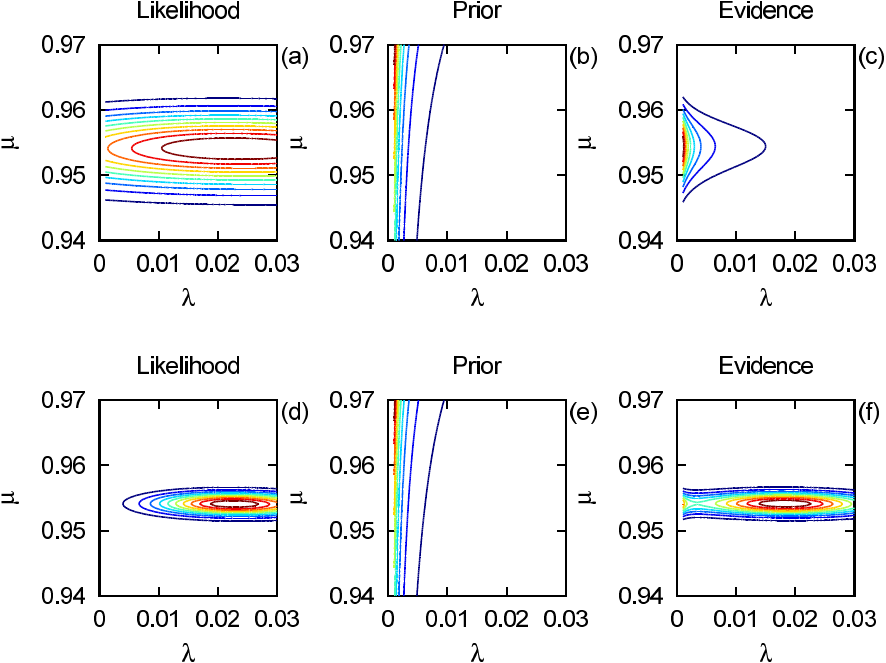}
\end{center}
\caption{Comparison of the likelihood, prior, and evidence densities in the analysis of data from Table~\ref{tab:D} in panels (a)--(c); for panels (d)--(f) the data is multiplied by a factor of 10.}
\label{fig:E}
\end{figure}

The evidence analysis of the same data is shown in Table~\ref{tab:E} for all five models.  A value of $\epsilon = 10^{-5}$ is selected, consistent with the amount of data $N_{GG} + N_{Gg} + N_{gg} = 1612$; in other words, we select a mathematical resolution slightly beyond that given by the measurement procedure which excludes the extreme boundary.  The location of the mode, when it exists, is displayed, as are the expected values of the parameters; for model $M_{\lambda,\mu}$ the global optimum of evidence is located on the boundary, and for model $M_{0,\mu}$ the mean value for $\mu$ is equal to $\mu_L$ as expected analytically.  The net evidence for each model $M$ is given in terms of its negative logarithm $\q{M}{\mbf{N}} \equiv - \log \mean{\p{\mbf{N}}{\mbf{m}}}{\mbf{m}}$ for parameter vector $\mbf{m}$ and data vector $\mbf{N} \equiv (N_{GG},N_{Gg},N_{gg})$.  The $Q$ value for each model, interpreted as the probability that the model describes the data, is determined from \beq
\bigq{M}{\mbf{N}} \equiv \exp (-\q{M}{\mbf{N}}) / \sum_M  \exp (-\q{M}{\mbf{N}}) \; ,
\eeq such that $\sum_M \bigq{M}{\mbf{N}} = 1$.  Of the two models supported by the data, that for Hardy--Weinberg equilibrium $M_{0,\mu}$ is assigned a probability close to 70\%.  A graphical comparison of the likelihood and evidence analysis is shown in Figure~\ref{fig:E} panels (a)--(c); in panels (d)--(f) we show the analysis of a hypothetical data set with 10 times as many observations per channel at the same ratios.  With that much data, the evidence is mostly around the likelihood peak, but a noticeable fraction is left along the manifold boundary.  For the amounts of data commonly found in observational studies, the prior can have a significant effect on the analysis.

Next let us look at some data from \citet{james-1983} as summarized by \citet{holzinger-popgen}, displayed in Table~\ref{tab:F}.  This time the data is broken down into that for subpopulations indexed by $k$ according to the geographic region of the observations.  Since none of the populations have only $N_{gg} > 0$, the model $M_0$ can be discarded immediately.  The practical question we are interested in is whether any single population is significantly different than the remainder.  To answer that question, the net evidence (expected likelihood) for the models applied to the entire population $\mbf{N}_0 \equiv \sum_k \mbf{N}_k$ is compared to the product of the evidence for the subdivision into $\mbf{N}_k$ and $\mbf{N}_{\sim k} \equiv \mbf{N}_0 - \mbf{N}_k$.  The results of this analysis are shown in Table~\ref{tab:G} using a value of $\epsilon = 10^{-5}$.  Values of 0 or 1 for the parameter mode appearing in the table are understood to be on the boundary given by $\epsilon$.

\begin{table}
\caption{Genotype observations from \citet{james-1983}}
\label{tab:F} 
\centering
\begin{tabular}{l *{12}{c}}
\hline
$k$ &  1 &  2 &  3 &  4 &  5 &  6 &  7 &  8 &  9 & 10 & 11 & 12 \\     
\hline                                                                 
$N_{GG}$ & 29 & 14 & 15 &  9 &  9 & 23 & 23 & 29 &  5 &  1 &  0 &  1 \\
$N_{Gg}$ &  0 &  3 &  2 &  0 &  0 &  5 &  3 &  3 &  0 &  0 &  1 &  0 \\
$N_{gg}$ &  0 &  3 &  3 &  0 &  0 &  2 &  4 &  1 &  0 &  0 &  0 &  0 \\
\hline
\end{tabular}
\end{table}

\begin{table}
\caption{Evidence analysis in terms of modes, means, and $q$ values of the data from Table~\ref{tab:F} for the entire population $\mbf{N}_0$, the subpopulations $\mbf{N}_k$ indexed by $k$, and the remainder populations $N_{\sim k}$ indexed by $k$}
\label{tab:G} 
\centering
\setlength{\tabcolsep}{4pt}
\begin{tabular}{*{13}{r}}
\hline
\multicolumn{1}{c}{$k$} & \multicolumn{1}{c}{$\lambda_{\lambda,\mu}$} & \multicolumn{1}{c}{$\mu_{\lambda,\mu}$} & \multicolumn{1}{c}{$\mu_{0,\mu}$} & \multicolumn{1}{c}{$\mu_{1,\mu}$} & \multicolumn{1}{c}{$\mean{\lambda}{\lambda,\mu}$} & \multicolumn{1}{c}{$\mean{\mu}{\lambda,\mu}$} & \multicolumn{1}{c}{$\mean{\mu}{0,\mu}$} & \multicolumn{1}{c}{$\mean{\mu}{1,\mu}$} & \multicolumn{1}{c}{$q_{\lambda,\mu}$} & \multicolumn{1}{c}{$q_{0,\mu}$} & \multicolumn{1}{c}{$q_{1,\mu}$} & \multicolumn{1}{c}{$q_1$} \\
\hline
 0&  0.55 &  0.89 &  0.89 &   NaN &  0.55 &  0.89 &  0.89 &   NaN & 110.05 & 125.93 &   Inf &   Inf \\ 
\hline                                                                                                
 1&  1.00 &  1.00 &  1.00 &  1.00 &  0.52 &  1.00 &  1.00 &  1.00 &  1.16 &  1.21 &  1.11 &  0.00 \\  
 2&  0.00 &  0.79 &  0.79 &   NaN &  0.45 &  0.77 &  0.78 &   NaN & 21.41 & 22.43 &   Inf &   Inf \\  
 3&  0.00 &  0.82 &  0.82 &   NaN &  0.62 &  0.80 &  0.80 &   NaN & 19.62 & 21.76 &   Inf &   Inf \\  
 4&  1.00 &  1.00 &  1.00 &  1.00 &  0.52 &  0.99 &  0.99 &  0.99 &  1.01 &  1.05 &  0.96 &  0.00 \\  
 5&  1.00 &  1.00 &  1.00 &  1.00 &  0.52 &  0.99 &  0.99 &  0.99 &  1.01 &  1.05 &  0.96 &  0.00 \\  
 6&  0.00 &  0.86 &  0.86 &   NaN &  0.19 &  0.85 &  0.85 &   NaN & 25.19 & 25.12 &   Inf &   Inf \\  
 7&  0.00 &  0.83 &  0.83 &   NaN &  0.63 &  0.82 &  0.82 &   NaN & 26.51 & 29.81 &   Inf &   Inf \\  
 8&  0.00 &  0.94 &  0.94 &   NaN &  0.20 &  0.92 &  0.92 &   NaN & 18.69 & 18.59 &   Inf &   Inf \\  
 9&  1.00 &  1.00 &  1.00 &  1.00 &  0.52 &  0.98 &  0.99 &  0.98 &  0.93 &  0.98 &  0.89 &  0.00 \\  
10&  1.00 &  1.00 &  1.00 &  1.00 &  0.52 &  0.93 &  0.95 &  0.91 &  0.74 &  0.78 &  0.69 &  0.00 \\  
11&  0.00 &  0.50 &   NaN &   NaN &  0.18 &  0.50 &  0.50 &   NaN &  3.85 &  3.14 &   Inf &   Inf \\  
12&  1.00 &  1.00 &  1.00 &  1.00 &  0.52 &  0.93 &  0.95 &  0.91 &  0.74 &  0.78 &  0.69 &  0.00 \\  
\hline                                                                                                
 1&  0.54 &  0.87 &  0.87 &   NaN &  0.53 &  0.86 &  0.86 &   NaN & 104.50 & 118.23 &   Inf &   Inf \\
 2&  0.54 &  0.90 &  0.90 &   NaN &  0.53 &  0.90 &  0.90 &   NaN & 91.97 & 104.33 &   Inf &   Inf  \\
 3&  0.52 &  0.90 &  0.90 &   NaN &  0.51 &  0.90 &  0.90 &   NaN & 94.29 & 105.81 &   Inf &   Inf  \\
 4&  0.55 &  0.88 &  0.88 &   NaN &  0.54 &  0.88 &  0.88 &   NaN & 108.43 & 123.69 &   Inf &   Inf \\
 5&  0.55 &  0.88 &  0.88 &   NaN &  0.54 &  0.88 &  0.88 &   NaN & 108.43 & 123.69 &   Inf &   Inf \\
 6&  0.61 &  0.90 &  0.89 &   NaN &  0.60 &  0.89 &  0.89 &   NaN & 88.24 & 103.50 &   Inf &   Inf  \\
 7&  0.51 &  0.90 &  0.90 &   NaN &  0.50 &  0.90 &  0.90 &   NaN & 87.66 & 97.79 &   Inf &   Inf   \\
 8&  0.58 &  0.88 &  0.88 &   NaN &  0.57 &  0.88 &  0.88 &   NaN & 94.90 & 109.60 &   Inf &   Inf  \\
 9&  0.55 &  0.89 &  0.88 &   NaN &  0.54 &  0.88 &  0.88 &   NaN & 109.16 & 124.70 &   Inf &   Inf \\
10&  0.55 &  0.89 &  0.89 &   NaN &  0.55 &  0.89 &  0.89 &   NaN & 109.87 & 125.69 &   Inf &   Inf \\
11&  0.57 &  0.89 &  0.89 &   NaN &  0.56 &  0.89 &  0.89 &   NaN & 107.59 & 124.32 &   Inf &   Inf \\
12&  0.55 &  0.89 &  0.89 &   NaN &  0.55 &  0.89 &  0.89 &   NaN & 109.87 & 125.69 &   Inf &   Inf \\
\hline
\end{tabular}
\end{table}

To identify which single population displays the most significant deviation from the remainder, for each $k$ the minimum $\q{M}{N_k}$ is added to the minimum $\q{M}{N_{\sim k}}$, then the minimum $\q{M}{N_0}$ is subtracted to yield the (negative) log evidence for the subdivision relative to the net population $\q{k}{0}$.  Those values are then exponentiated and normalized to yield the quality factors $\bigq{k}{0}$.  From Table~\ref{tab:G} one sees that $M_{\lambda,\mu}$ is the model best supported by the net population and all the remainder populations, but all four models can be supported by some of the subpopulations $N_k$.  In Table~\ref{tab:H} we display which model $M_k$ best fits population $N_k$ as well as the relative log evidence $\q{k}{0}$ and the quality factors $\bigq{k}{0}$ in units of percent.  Of the thirteen models under comparison, the most significant is the subdivision of the first population $k = 1$ from the remainder, whose $Q$ is close to 94\%.  The suppression of the recessive allele in that population would appear to be significant, while that for the other populations displaying only $N_{GG} > 0$ is less so.  A thorough analysis would consider all possible groupings of the subpopulations to determine the most statistically significant division of the net population from the given data.  A more thorough analysis would make use of knowledge of the geographical regions sampled to consider only those groupings of populations in physical contact.  The possibilities are endless and left as an exercise for the reader.

\begin{table}
\caption{Evidence comparison for the division of the data from Table~\ref{tab:F} into subpopulations of $N_k$ and $N_{\sim k}$ relative to the entire population $N_0$ with $\bigq{}{}$ stated in units of percent}
\label{tab:H} 
\centering
\setlength{\tabcolsep}{4pt}
\begin{tabular}{l *{13}{c}}
\hline
$k$ &  0 &  1 &  2 &  3 &  4 &  5 &  6 &  7 &  8 &  9 & 10 & 11 & 12 \\                                                                     
\hline                                                                                                                                      
$M_{k}$ & $\lambda,\mu$ & $1$ & $\lambda,\mu$ & $\lambda,\mu$ & $1$ & $1$ & $0,\mu$ & $\lambda,\mu$ & $0,\mu$ & $1$ & $1$ & $0,\mu$ & $1$ \\
$\q{k}{0}$ & 0.00 & -5.55 & 3.33 & 3.87 & -1.61 & -1.61 & 3.31 & 4.13 & 3.45 & -0.89 & -0.18 & 0.68 & -0.18 \\                              
$\bigq{k}{0}$ & 0.37 & 93.96 & 0.01 & 0.01 & 1.84 & 1.84 & 0.01 & 0.01 & 0.01 & 0.89 & 0.44 & 0.18 & 0.44 \\                                
\hline
\end{tabular}
\end{table}

\section{Discussion and conclusion}

Those who use Bayesian methods are often asked to explain the significance of the prior.  On its own, Bayes' theorem does not tell one how to assign the intrinsic probability density for the parameter manifold.  For that task, one must turn to some other maxim.  The principle of indifference is essentially a geometric argument that posits the existence of some coordinate mapping of the parameter manifold for which the information content is uniform.  That mapping might not be the one most convenient for the investigator, thus the appearance of the prior may be nonuniform in one's chosen coordinates.  The main effect of the prior is to prevent one from overestimating structure in the model not supported by imperfect data.  If the prior is neglected, one may unintentionally introduce a bias into one's results.

With respect to the beta distribution, use of the transformation group prior is implicit in its functional form.  In the absence of observations, what remains is the Haldane prior $\p{x}{} \propto x^{-1} (1-x)^{-1}$ expressing complete indifference to the value of an absolute probability.  If the observations $A$ and $B$ are restricted to integer counts of class membership, then the effect of the prior is to require an observation of each type of event before one is certain both types are present within the population; until both types have been observed, the evidence density is infinite on the boundary at either 0 or 1.  If one of each type has been observed, we are then certain that the production rate $x$ is between 0 and 1 with uniform distribution.  Further observations then refine that estimate until the likelihood and evidence modes converge in the limit of infinite data.

The transformation group approach leads one to specify $\q{A,B}{} = \log A + \log B$ as the logarithm of the unnormalized prior measure over the $(A,B)$ manifold. In the course of this project we investigated use of the entropic prior  $\p{A,B}{} \propto \exp (\mean{\q{x}{A,B}}{x \given A,B})$, where \beq
-\q{A,B}{} = \log \beta(A,B) + (A+B-2) \Lambda_1 (A+B) + (1-A) \Lambda_1 (A) + (1-B) \Lambda_1 (B)
\eeq is evaluated from the Shannon-Jaynes expression \citep{lazo-2269298}.  The entropic expression for the prior was discarded after finding in the context of the Balding-Nichols genotype analysis that it did not lead to a hierarchy of models.  The Jeffreys invariant prior, with logarithm \beq
-\q{A,B}{} = 2^{-1} \log \{ \Lambda_2 (A) \Lambda_2 (B) - [\Lambda_2 (A) + \Lambda_2 (B)] \Lambda_2 (A+B) \}
\eeq and proportional to the square root of the determinant of the Fisher matrix, likewise was considered.  Its prior density is very similar to that given by the transformation group, thus results based on that prior should be close to the results presented here.  Finally, the conjugate prior approach is discounted because there is no physical reason to suppose that the evidence and prior should be of the same algebraic form, mathematical convenience notwithstanding.  Note that the appearance of the beta function in the beta distribution results from the normalization over the axis $x \in [0,1]$; if the domain of $x$ is more restrictive, the expression for the normalization as a function of the parameters $A$ and $B$ is more complicated.  In that case, neither the entropic nor the Jeffreys prior is appropriate without severe modification, whereas the transformation group prior is unaltered.

Many investigators are troubled by the use of an improper prior, leading to an entire industry devoted to the generation of ever more complicated functions to be used as priors for statistical analysis of data.  One should think very carefully before deciding to employ any of those alternative strategies.  \citet{stern-2011symmetry} argues that good choices for the functional form of the probability densities used in a statistical model must be based on the natural symmetries and invariance properties of the quantities of interest.  The transformation group approach is based on the physical properties of the objects under consideration, with respect to the nature of the universe that we live in.  The prior it yields represents a measure of uniform information content over the parameter manifold.  The one dimensional improper transformation group priors are in fact all just different views of the uniform prior under a change of coordinates, $\int_0^1 dx / x (1-x) = \int_0^\infty dz / z = \int_\minfty^\infty du$ for $u = \log z$ and $z = x / (1-x)$.  The appearance of infinite densities on the boundary of the prior indicate where simpler models with fewer parameters exist; these models can be addressed by evaluating their Bayes factor relative to the model with the most complexity.

An outstanding issue when using the transformation group approach is the imposition of the finite cutoff $\epsilon$.  In the realm of physics, one argues that the measurement apparatus has a finite domain of resolution, from which a sensible value of $\epsilon$ may be derived.  Practically, one often sets $\epsilon$ to some value well beyond the expected domain of resolution with the understanding that one should check for boundary effects, and for many problems with well resolved parameters that is sufficient.  For counting experiments with a Poisson likelihood, the total number of observations is constrained by the patience of the investigator, thus it provides a finite limit to the resolution.  The sharp cutoff at $\epsilon$, though, is not appealing, when intuitively one expects a proper prior with finite normalization to have a smooth behavior.  Forthcoming in Part 2 is an extension of the approach presented here that incorporates depth of data in a manner that yields a smooth, normalizable prior function whose domain extends to the boundaries.

In summary, we have explored the relation between the Poisson and gamma distributions in the continuum with respect to the transformation group prior whose marginalization yields the beta distribution.  To impose normalization on the prior, we consider the limit on resolution of the parameters given by a finite set of observations.  Some examples of the approach are presented which display the flexibility of the beta distribution to model observational experiments.  Its study has a long history in the literature, and it continues to be quite useful in the modern day.

\appendix

\makeatletter   
 \renewcommand{\@seccntformat}[1]{APPENDIX~{\csname the#1\endcsname}.\hspace*{1em}}
\makeatother

\section{Normalization of the continuum Poisson distribution}
\label{sec:app}

In Section~\ref{sec:bpandg} we encountered an integral that could not be put into closed form analytically.  
In this appendix we present some heuristic arguments for its evaluation.  Let \beq
I(a) \equiv \int_0^\infty dA \, \dfrac{a^A}{\Gamma(A+1)} = \int_0^\infty dA \, \dfrac{a^A}{A \Gamma(A)}
\eeq represent the integral in question, and what we want to show is that $I(a) = e^a$.  Physically, the argument of the exponential function must carry no units, thus what we really mean by $e^a$ is \beq \label{eqn:expsum}
\exp (a / u_A) = \sum_{k = 0}^\infty \Delta_k (a / u_A)^k / k! = \sum_{k = 0}^\infty \Delta_k a^k / (k u_A)! = \sum_{A = 0}^\infty \Delta_A a^{A / u_A} / A! u_A \; ,
\eeq since $u_a = u_A$ and $\Delta_A = u_A \Delta_k$.  Similarly, $\Gamma(A)$ carries units of $u_a^A$, as can be seen from the Euler integral of the second kind $\Gamma(A) = \int_0^\infty da \, a^{A-1} e^{-a}$, thus $a^A / \Gamma(A)$ is a pure number, as is $I(a)$.  In taking the limit $\Delta_A \rightarrow 0$ of Eqn.~(\ref{eqn:expsum}), one must consider carefully the meaning of the denominator on the RHS.  When writing the factorial function as a product of descending integers, one typically stops at the factor 1; however, recalling that $0! = 1$, one sees that the factor $u_A$ completes the factorial so that the expressions $A! u_A \sim A \Gamma(A)$ carry the same units.  Since $\Delta_A$ does not appear on the LHS, taking the limit establishes the relation $I(a) = e^a$.

Without an antiderivative with respect to $A$ in hand for the densities $\p{A}{a}$ and $\p{a}{A}$, the most we can do analytically is investigate the properties of their integrals.  From the normalization of the joint density $\p{a,A}{} = (C_0 a A)^{-1} a^A / e^a \Gamma(A)$, whose units are carried by the first factor in parentheses, one can write \beq
1 = \int_0^\infty da \, \p{a}{} \int_0^\infty dA \, \p{A}{a} = \int_0^\infty da \, (C_0 a)^{-1} \int_0^\infty dA \, a^A / e^a \Gamma(A+1) \; ,
\eeq which implies that if $\dsub{a} \int_0^\infty dA \, \p{A}{a} = 0$, then $\int_0^\infty dA \, \p{A}{a} = 1$.  Since $\dsub{a} \p{A}{a} = (A / a - 1) \p{A}{a} = \p{a}{A} - \p{A}{a}$, one can say that \beq
\dsub{a} \int_0^\infty dA \, \p{A}{a} = \int_0^\infty dA \, \dsub{a} \p{A}{a} = \int_0^\infty dA \, \p{a}{A} - \int_0^\infty dA \, \p{A}{a} \; ,
\eeq whereby the establishment of $I(a) = e^a$ yields the normalizations $\int_0^\infty dA \, \p{A}{a} = 1$ and $\int_0^\infty dA \, \p{a}{A} = 1$, as required by the logical interpretation of Bayes' theorem, $\p{a,A}{} = \p{A}{a} \p{a}{} = \p{a}{A} \p{A}{}$.



\bibliographystyle{imsart-nameyear}
\bibliography{../stats,../solar,../plasma}

\end{document}